\documentclass[useAMS,usenatbib, usegraphicx]{mn2e}
\usepackage{times}
\title[X-ray spectral evolution of GX 349+2]
  {X-ray spectral evolution of low-mass X-ray binary GX 349+2}
\author[Agrawal et al.]
  {V. K.~Agrawal, 
  P.~Sreekumar  \\
   Space Astronomy \& Instrumentation Division, ISRO Satellite Centre Bangalore-17 \\ e-mail : vivekag@isac.ernet.in , pskumar@isac.ernet.in}
\date{Released 2002 Xxxxx XX}

\pagerange{\pageref{firstpage}--\pageref{lastpage}} \pubyear{2002}

\def\LaTeX{L\kern-.36em\raise.3ex\hbox{a}\kern-.15em
    T\kern-.1667em\lower.7ex\hbox{E}\kern-.125emX}

\begin{document}

\label{firstpage}

\maketitle

\begin{abstract}
We present the results of a systematic investigation of spectral
evolution in the Z source GX 349+2, using data obtained during 1998
with Proportional Counter Array (PCA) on-board the RXTE satellite. The
source traced a extended normal branch (NB) and flaring branch (FB)
in the color-color diagram (CD) and hardness-intensity diagram (HID)
during these observations. The spectra at different positions of
Z-track were best fitted by a model consisting of a disk blackbody
and a comptonized spectrum. A broad (Gaussian) iron line at $\sim
6.7$ keV is also required to improve the fit. The spectral parameters
showed a systematic and significant variation with the position along
the Z-track. The evolution in spectral parameters is discussed in the
view of increasing mass accretion rate scenario, proposed to explain
the motion of Z sources in the CD and HID.
\end{abstract}


\section{Introduction} 
Low Mass X-ray binaries (LMXBs) containing neutron stars are broadly
divided into two classes: Z and atoll sources \citep{hga}. This
classification is based on the shape of the path traced by these sources
in the X-ray color-color diagram (CD) or hardness-intensity diagram
(HID).  Z-sources trace out a Z-type pattern in CD and HID, and have
luminosities close to Eddington limit (L$_{Edd})$. A Z-track consists
of three parts, horizontal branch (HB), normal branch (NB) and flaring
branch (FB).  Atoll sources trace out a fragmented shape in CD. They have
luminosities 0.01-0.1 L$_{Edd}$ and probably, have weaker magnetic fields
than Z-sources \citep{vaa, ps}. It is widely believed that in Z-sources
the inferred mass accretion rate increases along the Z-track from HB to
FB \citep{hgb}. Similarly, in atoll sources inferred mass accretion rate
increases along the fragmented path from the island state to upper part
of the banana branch. \\ 
X-ray spectra of LMXBs can be described as the sum of soft and hard
components.  Two different approaches have been used to model these
components. In the first case, the soft component is modeled by a single
temperature blackbody corresponding to the hot surface of neutron star and
a hard component corresponding to comptonized emission from the inner disk
or boundary layer \citep{wh, ws, dia, dib}.  In the second approach, the
soft component is described by a multicolor disk blackbody from accretion
disk and the hard component is modeled by comptonized emission \citep{mta,
mtb, dic}.  Comparison of these models with observed data have shown
that both the approaches require an additional broad iron $K_\alpha$
line, arising from the ionized disk or from the accretion disk corona
(ADC).  Atoll sources exhibit two distinct spectral states; namely hard
and soft.  In hard state their spectrum extends upto energies $_\sim ^>$
100 keV \citep{ba}.  The spectrum of Z sources is much softer, with cutoff
energies of a few keV. In addition, a hard tail has been occasionally
seen in the spectrum of the 5 Z sources (GX 5-1:\citealt{as}; GX 17+2:
\citealt{dia}; GX 349+2: \citealt{dib}; Sco X-1: \citealt{df}; Cyg X-2:
\citealt{ff}; \citealt{dic}). 

 UV/optical observations \citep{vr} and correlated spectral and timing
 study of Z sources suggest \citep{vaa} that accretion rate increases
 along the Z-track from HB to FB.  Modeling of X-ray spectra of Z sources
 also indicate that their motion in the CD is caused by variations in the
 inferred mass accretion rate \citep{ps}. However, other possibilities
 also exist \citep{vab}. Thus, investigations of spectral evolution along
 the CD may provide an important clue to understand this scenario.  The
 evolution of X-ray spectra of several Z sources have been investigated
 using BeppoSAX ( GX 17+2: \citealt{dia}; GX 349+2: \citealt{dib}; Cyg
 X-2: \citealt{dic}).  In GX 349+2,  no appreciable change was observed in
 spectral parameters as source moved from NB to FB. Only, seed photon and
 electron temperature were slightly higher in FB than NB. Also, in GX 17+2
 spectral parameters  did not vary significantly from HB to NB. X-ray
 spectra of Cyg X-2 showed an interesting evolution along the Z-track
 \citep{dic}. It was found that as the source Cyg X-2 made a transition
 from HB to NB the inner disk temperature increased and at the same
 time inner rim of accretion disk moved inward. Also,  optical depth of
 comptonized component systematically decreased and electron temperature
 increased as the source moved from HB to NB. The authors suggest that
 hardening of the soft blackbody component and softening of the hard comptonized
 emission are probably responsible for HB to NB transition.
 \\
    In this paper, we present the most detailed investigation of spectral evolution in GX  349+2 using extensive data set obtained with RXTE. The source traced the most complete CD during our observations (\citealt{zw}; \citealt{ag}; \citealt{opm}), providing an opportunity to study the spectral evolution in detail.

\section{Observation}
We analyzed public archival data obtained with PCA instrument
on-board RXTE satellite. The data were collected during January  and
September-October, 1998. The total observation time with good data from
PCA is $\sim$~220 ks. The PCA consists of five identical proportional
counter units with total effective area of $\sim$ 7000 cm$^2$ in the
energy range 2-60 keV \citep{jh}. 
\begin{figure}
\includegraphics[width=65mm, angle=-90]{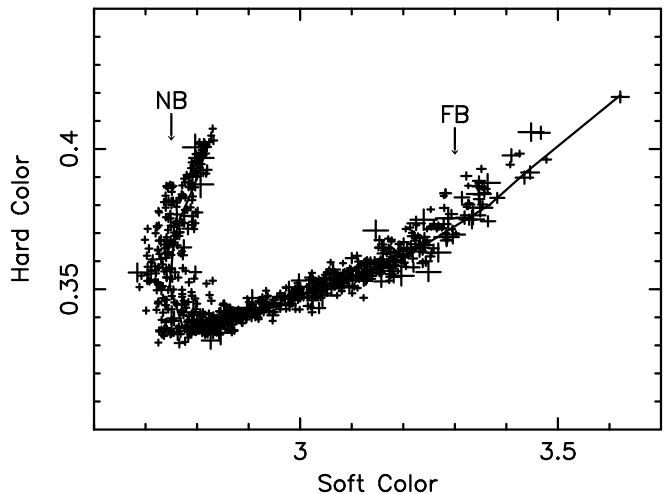}
 \caption{Color-color diagram of GX 349+2 for the January and September-October 1998 data sets. Soft color is count-rate-ratio in 3.5-6.4 keV and 2.0-3.5 keV energy ranges and hard color is that in energy bands 6.4-9.7 keV and 9.7-16.0 keV. The solid line in this figure represents an average Z-track passing through normal points. Each points is derived from a time bin size of 256s. }
 \label{sample-figure}
\end{figure}

\section{Color-Color diagram}
The PCA standard-2 mode data, accumulated every 16s in the energy
range 2-60 keV were used to create X-ray color-color diagram (CD) and
Hardness-Intensity diagram (HID). The soft color is defined as the ratio
of count rates in 3.5-6.4 keV and 2.0-3.5 keV energy ranges and hard color
is defined as that in 9.7-16.0 keV and 6.4-9.7 keV energy ranges. The
`rank number' or `$S_z$' parameterization technique \citep[][and
reference there in]{disw} was used to define the position of the
source in CD.  We selected the normal points in the CD in such a way
that they form a smooth curve. The color-color points in the CD are
projected onto this curve. The $S_z$ parameter for each projected point
was calculated by measuring their distance from NB/FB vertex. NB/FB
vertex ($S_z$ = 2) and end-point of the FB ($S_z$ = 3) were taken as
reference points and rest of the Z curve was normalized according to the
length of FB. The CD obtained from January and September-October
observations of this source is shown in figure 1. The solid curve in
figure 1 represents the approximate Z-track.

\begin{figure}
\includegraphics[width=100mm,height=80mm,angle=-90]{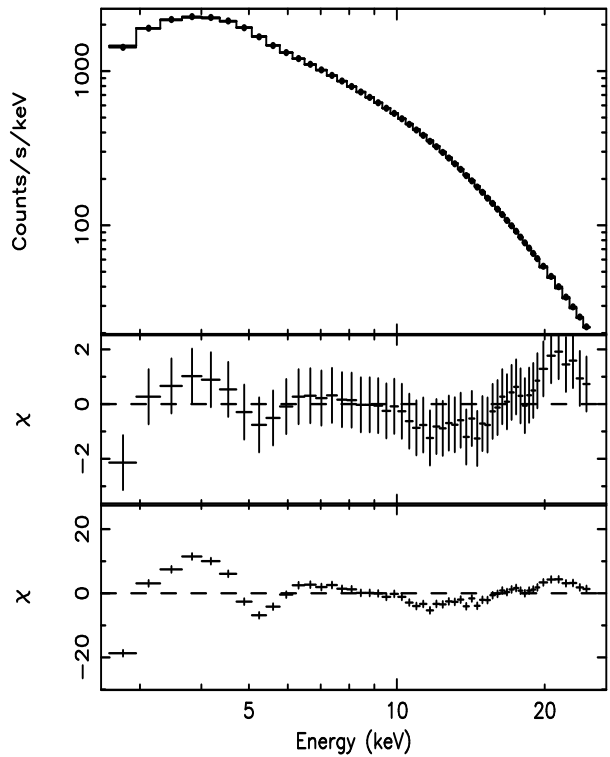}
 \caption{Count-rate spectrum and folded model ({ \bf power-law}) for the Crab is shown in panel 1. The residuals in the unit of $\sigma$ are shown in panel 2 (a systematic error of 1 \% is included) and panel 3 (without an inclusion of systematic error). The filled circle represent the observed count rate spectrum.}
 \label{sample-figure}
\end{figure}

\section{Spectral analysis}
The standard-2 mode PCA data were used to create the spectra in
the energy range 2.5-25.0 keV. {\bf The new {\it cmbrightvle} model file
``pca\_bkgd\_cmbrightvle\_e3v20020201'' for epoch 3 was used to calculate
the PCA background.} We used the observation (from obsid 30133-01-02-00
to  30133-01-02-08) of Crab Nebula  to derive the level of uncertainty
in the PCA response matrix. Fitting the Crab spectrum (in the energy
range 2.5-25.0 keV) to a power-law with photon-index ($\Gamma$) =
2.185$\pm$0.005 provided a reduced $\chi^2$ ($\chi^2_\nu = \chi^2/d.o.f)$
of 911/51. But, an inclusion of 1\% systematic error reduced the
value of $\chi ^2_\nu$ to 41/51. Hence, a systematic error of 1\%
was added to all the spectra to take into account the uncertainty in the
PCA response matrix. In figure 2, the effect of PCA systematics on
the Crab spectrum has been shown.
\begin{figure}
\includegraphics[width=100mm,height=80mm,angle=-90]{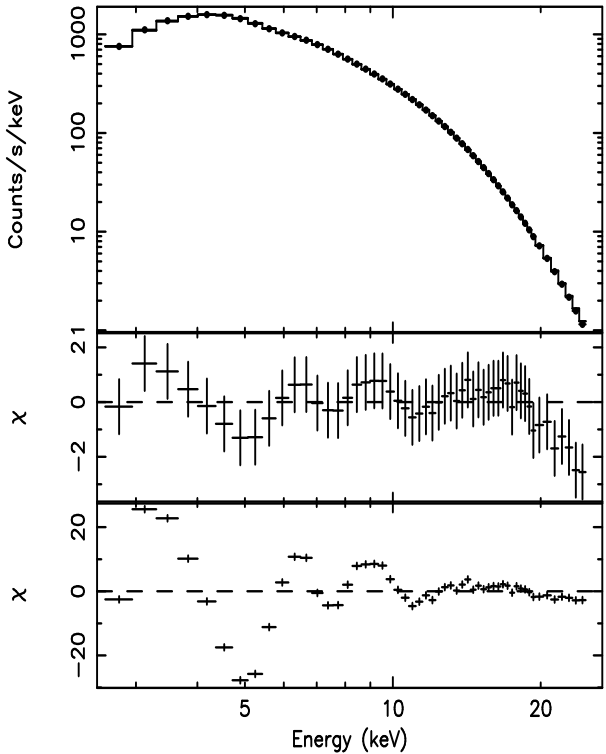}
 \caption{Count-rate spectrum with folded model ({ \bf diskbb+comptt+gauss+pl}) for the interval NB2 is shown in panel 1. The residuals in the unit of $\sigma$ are shown in panel 2 (a systematic error of 1 \% is included) and panel 3 (without an inclusion of systematic error). The filled circles represent the observed count rate spectrum.}
 \label{sample-figure}
\end{figure}

  We divided the NB into 5 intervals and FB into 8 intervals to
  make an extensive investigation of  spectral evolution in the source
  GX 349+2 along the Z-track. The source and background spectra within
  each interval were added separately to form a cumulative  source
  and background spectrum. The background subtracted energy spectrum
  in the 2-25 keV range were fitted using a  two-component model. The
  multi-temperature blackbody ({\bf diskbb} in XSPEC, Mitsuda et al. 1984)
  or a single temperature blackbody ({\bf bbody } in XSPEC) was used to
  describe the soft component. To model the hard component, comptonization
  models {\bf compst} \citep{sra} and {\bf comptt} \citep{tl} of XSPEC
  were used. We find that {\bf diskbb+comptt} model provided a better
  fit ($\chi^2_\nu = 133/48)$ to the spectrum at upper NB (NB1)
  compared to {\bf bbody+comptt} model ($\chi^2_\nu = 158/48)$.
  The {\bf diskbb+compst} model provides a poor fit to the spectrum for
  interval NB1 ($\chi^2_\nu \sim$ 643/49 ).  It is to be noted that,
  in our case the temperature of diskbb component is nearly coinciding
  with lower energy range of the detector and hence {\bf diskbb} and {\bf
  bbody} model cannot be discriminated clearly.  The analysis of X-ray
  bursts from GX 17+2 has suggested that the  soft component comes from
  the optically thick accretion disk \citep{kkc}. Therefore, we
  chose {\bf diskbb+comptt} model to  describe the 2.5-25.0 keV spectra
  at various parts of Z-track. A Gaussian emission line at $\sim$
  6.7 keV is required to improve the fit to the spectra for all the
  positions on the Z-track. Addition of this line decreased $\chi ^2_\nu$
  from 133/48 to 30.7/45 for the spectrum of NB1 ( probability of chance
  improvement of fit is $\sim$ $2.3\times10^{-14}$). The count rate
  spectrum for the source GX 349+2 (during interval NB2) in the energy
  range 2.5-25 keV and best-fit to it are shown in the panel 1 of figure
  3. The residuals in the unit of sigma, with and without inclusion of
  systematic error are shown in panel 2 and 3 respectively.  
\begin{figure}
\includegraphics[width=100mm,height=80mm,angle=-90]{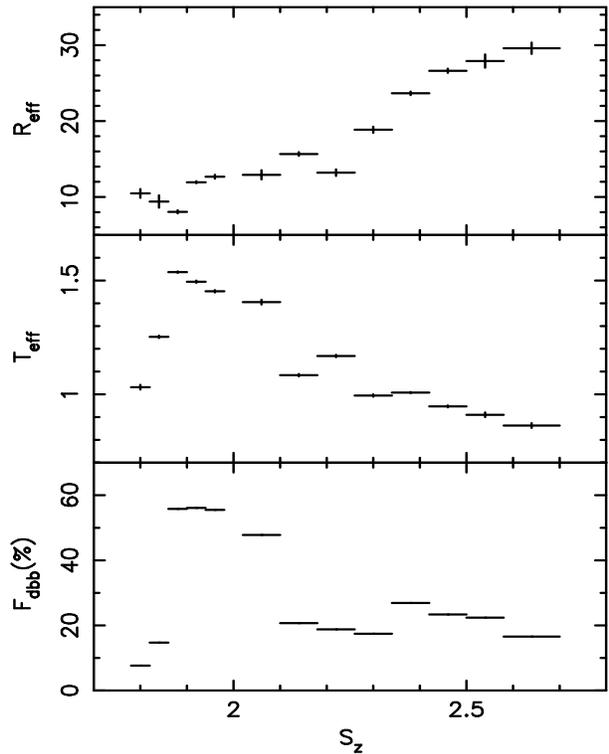}
 \caption{Variation of the best-fit {\bf diskbb} parameters as a function of position (in terms of $S_z$ parameter) along the Z-track. The inner disk temperature $T_{eff}$ is given in keV and the inner disk radius $R_{eff}$ is given in km. $F_{dbb}$ (in \%) gives percentage contribution of {\bf diskbb} flux to the total flux (in the energy range 2.5-25.0 keV). }
 \label{sample-figure}
\end{figure}


\section{Results and Discussion}
The values of best fit spectral
parameters are given in Table 1 (for NB) and Table 2 (for FB). The
diskbb model represents the soft emission expected from the optically
thick accretion disk and provides two important disk parameters, color
temperature ($T_{col})$ of inner disk and inner disk radius ($R_{in})$.
The inner disk radius $R_{in}$  is expressed by:
\begin{equation}
\left(\frac{R_{in}}{km}\right) = \frac{\sqrt{K}\times\left(\frac{D}{10~kpc}\right)}{\sqrt{\cos\theta}}
\end{equation}
In this expression K is the disk blackbody normalization, D is the
distance to the source in kpc, and $\theta$ is the inclination of the
accretion disk. Distance of the source GX 349+2 is uncertain
and probably lies in the range 5-10 kpc \citep{cp, chsw, vpm} ).  Optical
observation suggests that if GX 349+2 is a LMXB ( mass of companion  $<
2 M_\odot)$ then inclination should be  $\theta > 25^\circ$ for the
neutron star of mass $ < 2 M_\odot$ \citep{wm}.   X-ray properties of
Z-sources suggest that there are two different classes of Z-sources
\citep{kka, kkb} : Cyg-like ( Cyg X-2, GX 5-1, Sco X-1) and Sco-like
sources (Sco X-1, GX 349+2, GX 17+2).  One of the best studied Cyg-like
source Cyg X-2 has an inclination $\theta \sim 60^\circ $\citep{jk}.
As, it has been proposed that Sco-like sources are being viewed at lower
inclination than Cyg-like sources \citep{kka, kkb},  for the source GX
349+2 inclination should be $ 25^\circ <\theta < 60^\circ$. Therefore
we assume a mean inclination $\theta=45^\circ$ and distance D = 7.5 kpc
for GX 349+2 to derive its inner disk radius. It is to be noted that a different choice of inclination and distance in the range specified above will
scale the inner disk radius by a factor of 0.59 to 1.58.
However, as the inner disk radius for all the positions on the Z-track
are scaled by the same amount, an overall trend of the change in the
inner disk radius will not be affected.

In the multicolor disk model the local emission is described by blackbody
spectrum \citep{ss}. However, it is expected that in the inner region
of the disk, opacity due to electron scattering dominates over that due
to absorption processes.  In this case the local spectrum can be
significantly modified by comptonization and can be approximated by,
\begin{equation}
I_\nu = \frac{1}{f^4} B_\nu (T_{col})
\end{equation}

where, $f = \left(\frac{T_{col}}{T_{eff}}\right)$ is the spectral
hardening factor , $T_{col}$ is the color temperature of inner disk,
$T_{eff}$ is the effective temperature of inner disk, $\nu$ is the
frequency of radiation, $B_\nu (T_{col})$ is the Planck function
\citep{ebk}.  The effective radius and temperature of inner disk is
given by, 
\begin{equation}
R_{eff} = f^2 R_{in}
\end{equation}
\begin{equation}
T_{eff} = \frac{T_{col}}{f}
\end{equation}

By solving the vertical structure and radiative transfer equation of the
disk, \citet{sht} showed that for viscosity parameter $\alpha \sim 0.1$,
luminosity close to Eddington limit and mass of the compact object 1.4-10
$M_\odot$ the soft emission from disk can be described by equation 2
with spectral hardening factor f=1.9$\pm$0.1.

 The variation in the parameters of soft component as a function of $S_z$
 parameter is shown in figure 4. Our study revealed that $T_{eff}$
 increased from $\sim$ 1 keV to $\sim$ 1.5 keV as source moved from
 upper to middle part of NB, i.e in the direction of increasing $S_z$.
 Increase in $T_{eff}$ can be interpreted in terms of inward motion
 of accretion disk with increasing accretion rate. In fact inner disk
 radius $R_{eff}$ was found to decrease from upper NB ( $\sim$ 10.5 km)
 to middle NB($\sim$ 8.0 km).   As source moved further from middle NB
 to upper FB,  $T_{eff}$ decreased from $\sim$~1.5 keV to $\sim$~ 0.86
 keV. At the same time $R_{eff}$ increased from $\sim$~12 km to $\sim$~
 30 km. This suggests that inner rim of accretion disk moves outward
  again with increasing mass accretion rate.  

As mentioned earlier, {\bf diskbb} radiation is emitted by  optically
thick matter, which is in Keplerian flow. Since the radiation drag
force exerted by radiation coming from the neutron star can remove
significant angular momentum from the gas in Keplerian flow, radial
flow starts a few stellar radii above the neutron star surface.  The inner
edge of the accretion disk can be taken as the point at which disk-flow
ends and radial flow starts. Note that the intensity of radiation will
be attenuated exponentially ($\exp^{-\tau_{r}}$, where $\tau_r$ is the
radial optical depth) as it passes through disk.  Therefore radiation
drag cannot remove any angular momentum beyond a certain distance
(R$_{aml})$ from the central compact object and supersonic radial flow
is possible only within radius $R_{aml}$ \citep{ml, mlb}. This radius is
obtained by setting $\tau_r$ = 5 (since most of the radiation  will be
attenuated beyond this optical depth). Since radial flow probably begins
at the inner edge of the accretion disk, R$_{aml}$ should be very close
to $R_{eff}$, i.e, $R_{aml} \approx R_{eff}$.  Radial optical depth of
accreting matter is given by equation 5 of \cite{ml}, which is
\begin{equation}
\tau_r = \frac{\sigma_T \dot M}{4\pi m_p}\int_R^r \frac{[1-v^{\hat{r}}(r')]^{-1}}{\gamma_r (r') \gamma(r')} \frac{(1-2M/r')^{-1}}{r' h(r') v^{\hat{r}} (r')}~dr' .
\end{equation}
In this expression, $\dot M$ is the mass accretion rate through
the inner disk, $\sigma_T$ is the Thomson scattering cross section,
$v^{\hat{r}} (r')$ is the radial velocity, $h(r')$ is the half-thickness
of disk, $\gamma_r = [1-(v^{\hat{r}})^2]^{-1/2}$, $\gamma =
[1-{v^{\hat{\phi}}}^2-{v^{\hat{r}}}^2]^{-1/2}$, where $v^\phi$ is azimuthal
velocity of accretion flow. Note that in all the expressions of this
section we have used units such that c =  G = 1, where c is the speed of
light and G is gravitational constant.  By substituting $\tau_r = 5$ and
solving equation 1 for radius (see equation 6 of \cite{ml}) one can get

\begin{equation}
R_{eff} \approx R+5 \left( \frac{\dot M}{0.01~\dot M_E} \right )^{-1} \left (\frac{R}{10~km} \right ) \left (\frac{v^{\hat{r}}}{0.1c} \right ) \left(\frac{h/R}{10^{-2}} \right ) ~~km.
\end{equation}

From equation 6, it is clear that inner disk radius will decrease with
increasing accretion rate. This explains decrease of inner disk radius
seen from upper NB to middle NB. However equation 6 cannot explain
the increase of inner disk radius from lower NB to upper part of FB.
In this part of CD (from lower NB to upper FB) the source is supposed
to be  super-Eddington and hence gravitational pressure cannot support
the radiation pressure. This will lead to an outflow in this part of
the Z-track. It should be noted that equation 5 and 6 do not take into
account the presence of outflow. The presence of outflow can remove
the matter from radial inflow and can dump it into a hot central corona
above the inner disk. This  will lead to a reduction in the density of
radial flow and hence, in radial optical depth. A decrease in the
optical depth of radial flow will cause radiation to penetrate further
inside the disk and hence the inner rim of the disk will shift outward
(i.e, $R_{eff}$ will increase).  Also, if a fraction of matter from
radial inflow is pumped in to a hot corona then the optical depth of
this corona will increase. In fact optical depth of the corona is found
to increase along the FB where most likely $\dot M > \dot M_{Edd}$.  

The contribution of the soft component (F$_{dbb}$) to the total flux increases
from $\sim$ 8\% to 56\% as the source moves from upper to lower NB. Again from
lower part of NB to upper FB, the soft emission decreased from $\sim$56\%
to 17\%. The variation in the soft component can be explained in terms of
variations in the geometry of inner accretion disk with mass accretion
rate. An inward motion of disk (from upper to middle NB) increased the
area, emitting the soft component. As a result soft disk emission  also
increased. Similarly, an outward motion of disk (from lower NB to upper
FB) caused a decrease in  the soft disk flux. 
\begin{figure}
\includegraphics[width=100mm,height=80mm,angle=-90]{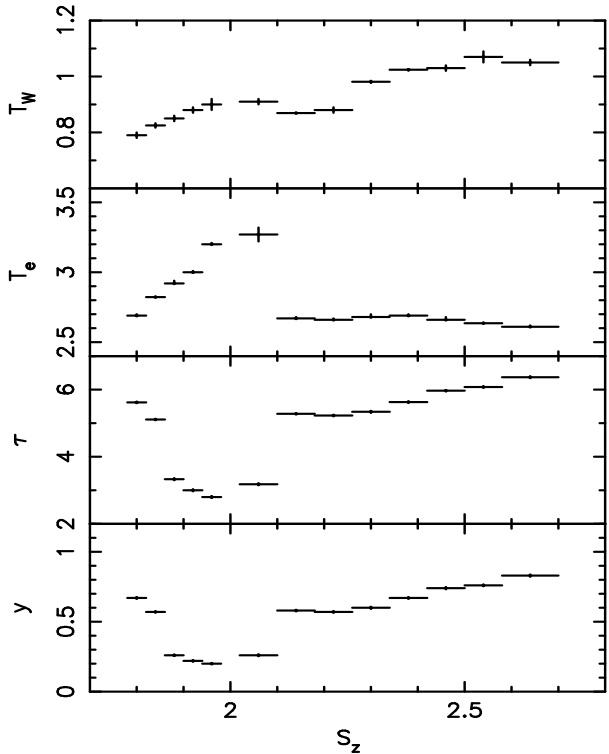}
 \caption{Variation in the best-fit {\bf comptt} parameters as a function of position (in terms of $S_z$ parameter) along the Z-track. Electron temperature $T_e$ and seed photon temperature $T_W$ are given in the unit of keV.}
 \label{sample-figure}
\end{figure}

The comptonized spectrum is produced by upscattering of soft photons
(described by the Wien spectrum at $\sim$~1 keV) by hot electrons in
the central hot corona or boundary layer. Our analysis showed that
parameters of the comptonized spectrum evolved significantly as the
source moved along the Z-track (see  figure 5).  The electron temperature
$T_e$ of comptonizing plasma cloud increased from $\sim$ 2.7 keV to 3.2
keV along the NB. At the same time optical depth $\tau$ of the plasma
cloud decreased from $\sim$ 5.6 to 2.8. Probably, increased heating of
the central corona caused by increase in the incident flux reduces the
density of corona and hence, decrease $\tau$. A decrease in $\tau$ leads
to a reduction in relative energy gain of the comptonization process,
given by comptonization parameter $y=\frac{4T_e}{m_e c^2}\tau^2$ (where
$m_e c^2$ is the rest mass of the electron), from upper to lower part of
NB. Other possibility is that a slow collapse of the central corona to
a geometrically thin accretion disk, leaving low density plasma cloud
above the accretion disk \citep{dic} may lead to a decrease
in optical depth. In this case same amount of matter will be accreted
on a smaller area (a strip around neutron star).  This will cause an
increase in seed photon temperature. In fact seed photon temperature $T_W$
increased slowly from 0.79$\pm$~0.01 keV to 0.90$\pm$~0.02 keV along the
NB. However, it is not clear that why a central corona should collapse
to a geometrically thin disk. As source moved further from NB/FB vertex
to upper FB, $T_e$ decreased from $\sim$ 3.3 keV to 2.6 keV and $\tau$
increased from $\sim$ 3.2 to 6.4. In the FB, the radiation pressure
is expected to be very high and dominant over gravitational pressure.
Therefore, probably due to effect of radiation pressure, an outflow
begins near the inner part of disk, creating a density enhancement in
the hot corona above the inner disk and hence causing optical depth to
increase along the FB. The comptonization parameter $ y$ also increases
along the FB due to increase in optical depth. 

The results  of \citet{dib} on the source GX 349+2 from
BeppoSAX  data are different from our results. In their work, the whole
Z-track was divided into two intervals; NB and FB. The  spectra  of FB and
NB were best fitted with {\bf bbody+comptt+power-law+Gaussian-lines}. As
the source GX 349+2 moved from NB to FB temperature of the blackbody
component slightly increased from 0.51$\pm$0.01 keV to 0.59$\pm$0.02
keV. At the same time  optical depth of the comptonizing cloud decreased
from 11.7$\pm$0.4 to 10.5$\pm$0.5 and electron temperature increased from
2.65$\pm$0.05 keV  to 2.95$\pm$0.07 keV.  The comptonization parameter $y$
was $\sim$ 2.84 in the NB and $\sim$ 2.54 in the FB.  Our results show
that diskbb component is harder in the NB ($T_{eff} = 2.75\pm0.01$) than
FB ($T_{eff} = 1.95\pm0.01$) and comptonized emission in the NB is soft
($y \sim 0.38$) compared to that in the FB ($ y \sim 0.63)$. {\bf There may
be two possible reasons behind the differences seen between our results
and those of \citet{dib}.  First, they have used a model
which is different from that used in our work. Second, they have used
different energy  ranges to model the NB (0.1-200 keV) and FB (1.8-200
keV) spectra as LECS data were not available during the flaring state.} 

We calculated the radius of seed photon emitting region ($R_W$) using
expression, $R_W = 3\times 10^4 D {\sqrt{\frac{F_{bol}}{1+y}}}/{(T_W)^2}$
as given by \citet{ij}. Here, D is the distance in kpc, $F_{bol}$ is the
unabsorbed comptonized flux in $ergs ~ s^{-1} cm^2$, $y$ is comptonization
parameter and $T_W$ is seed photon temperature in keV. The Wien radius
$R_W$ was $\sim$ 30-48 km in the NB and $\sim$ 30-40 km in the FB. These
values are considerably larger than that reported previously \citep{dib}.

We found a strong broad iron line at  energies $\sim$ 6.7 keV, with width
$\sim$ 0.7-1.4 keV.  It is to be noted that iron line was {\bf stronger} in the
FB (eq. width = 154-295) than NB ( eq. width = 97-210). This behavior
of GX 349+2 is just opposite to that exhibited by it during earlier
observations \citep{dib}. We also note that in our case observed iron
line is much stronger.

Even though X-ray intensity decreases or remains almost constant
along the NB, it is generally believed that accretion rate increases
along the Z-track.  This arguments mostly relies upon two observational
facts. First, UV and optical flux increases along the Z-track from HB to
FB \citep{vr,vp}. UV and optical fluxes are thought to be better indicator
of accretion rate than the X-ray flux \citep{hgb}. Second, frequency of
Horizontal-Branch-Oscillation (HBO) and kHz quasi-periodic-oscillation
(QPO) increases from HB to NB \citep[][and reference there in]{vab}. In
almost all the models, frequency of these two oscillations is directly
proportional to accretion rate \citep{lf, psb, ml, sl}. However, there
are a few recent results which indicate that increasing accretion
rate scenario may not be the true story \citep{hj} and motion along
Z-track is caused by variation in the inner disk radius \citep{vab}.
Our systematic investigation of spectral evolution showed that the
motion of the source along the Z-track can be explained by variation
in two emission components; soft disk emission and hard comptonized
spectrum. The motion of the source along the NB (i.e softening of
spectra) is caused by an increase in disk contribution and a decrease
in degree of comptonization. Similarly, a decrease in the contribution
of the soft component and an increase in Compton parameter $y$
seems to be responsible for the motion of the source along the FB. As
discussed earlier, the above changes in the soft component and in the
comptonized spectrum are expected if accretion rate is increasing along
the Z-track. Hence, our results seems to be in the favor of  the scenario
in which motion of the source along the Z-track is caused by variations
in the accretion rate.
\section{Conclusion}
In this paper, we have carried out a detailed study of spectral evolution
in the source GX 349+2 using RXTE data. During our observations
the source is found both in the NB and FB. The continuum spectra at
different positions of Z-track are well described by a disk blackbody
and a comptonized component. We explain the evolution of parameters of both
soft and hard components in terms of variations in the mass accretion rate
along the Z-track. This study suggests that a systematic increase in the
strength of soft component and a decrease in comptonization parameter
$ y$ are responsible for the motion of the source along the NB. Similarly,
a decrease in soft disk emission and increase in the $ y$ parameter
may cause the source to move along the FB. 

\section*{acknowledgments}
This work has made use of High Energy Astrophysics Science Archive Research
Center (HEASARC) facility at NASA-Goddard Space Flight Center. We are thankful 
to the anonymous referee for his useful suggestions.
  
\newpage
 \vspace{3.5cm}

\begin{center}
\begin{table*}
\begin{minipage}{120mm}
\caption{Results of diskbb+comptt+gauss fit to the NB spectra of GX 349+2 in the 2.5-25.0 keV energy range. For each spectrum the range of $S_z$ is given in the bracket. Errors on the best fit parameters are calculated by using $\Delta\chi^2=1$ (68\% confidence)}
\begin{tabular}{cccccc} \hline
 Para. &NB1 & NB2 & NB3 & NB4&NB5\\ 
  & (1.80$\pm$0.02) & (1.84$\pm$0.02) & (1.88$\pm$0.02) & (1.92$\pm$0.02)&(1.96$\pm$0.02) \\ \hline
$T_W$ (keV)  & 0.79$\pm$0.01  & 0.825$\pm$0.008    &  0.85$\pm$0.01 & 0.88$\pm$0.01 &0.90$\pm$0.02 \\
$T_e$ (keV)  & 2.69$^{+0.01}_{-0.004}$& 2.822$\pm$0.004    &  2.92$\pm$0.02& 3.00$\pm$0.01 &3.20$\pm$0.01\\
$\tau$  & 5.62$\pm$0.02 & 5.11$\pm$0.01    &  3.33$\pm$0.03 & 3.00$\pm$0.03 &2.80$\pm$0.03\\
$R_W$ (km)& 48.35$\pm$2.2 & 44.07$\pm$1.41 & 22.71$\pm$1.1 &29.97$\pm$1.35&29.28$\pm$1.05 \\ 
y & 0.67$\pm$0.006 &0.57$\pm$0.003 & 0.26$\pm$0.005 & 0.22$\pm$0.005 & 0.20$\pm$0.004 \\
$T_{eff}$ (keV)  & 1.03$\pm$0.01 & 1.25$\pm$0.005    &  1.54$\pm$0.003 & 1.49$\pm 0.005$ &1.45$\pm$0.005\\
$R_{eff}$ (km) &  10.47$\pm$0.58 & 9.40$\pm$0.81 & 8.04$\pm$0.21& 11.94$\pm$0.14& 12.67$\pm$0.26 \\  
$F_{dbb}(\%)$ & 7.64 & 14.7& 55.80 & 56.10 &55.51\\
$E_l$ (keV)   & 6.76$\pm$0.2 & 6.89$^{+0.27}_{-0.16}$ & 6.80$\pm$0.2 & 6.78$\pm$0.16 &6.80$\pm$0.14 \\
$\sigma_l$ (keV) &0.77$^{+0.25}_{-0.14}$ & 0.61$\pm$0.34    & 0.72$\pm$0.15 & 0.73$^{+0.26}_{-0.11}$& 0.68$^{+0.23}_{-0.11}$\\
Eq. Width (eV)& 155 & 97 & 142 & 184 & 210 \\
$\chi^2 (d.o.f)$& 30.7 (45) & 31.0 (45) & 32.8 (45) & 31.0 (45) & 36.0 (45) \\
\hline
\end{tabular}
\end{minipage}
\end{table*}

\begin{table*}
\tiny
\begin{minipage}{120mm}
\caption{Results of diskbb+comptt+gauss fit to the FB spectra of GX 349+2 in the 2.5-25.0 keV energy range. For each spectrum the range of $S_z$ is given in the bracket. Errors on the best fit parameters are calculated by using $\Delta\chi^2=1$ (68\% confidence)}
\begin{tabular}{ccccccccc} \hline
 Para. & FB1 & FB2& FB3 & FB4 & FB5 & FB6 & FB7 & FB8 \\
       &( 2.06$\pm$0.04) & (2.14$\pm$0.04)&(2.22$\pm$0.04)& (2.30$\pm$0.04) & (2.38$\pm$0.04) & (2.46$\pm$0.04) & (2.54$\pm$0.04) & (2.64$\pm$0.06)\\ \hline
$T_W$ (keV) & 0.91$\pm$0.01& 0.869$\pm$0.001&0.88$\pm$0.01 &0.981$\pm$0.004&1.024$\pm$0.003 & 1.03$\pm$0.01& 1.07$\pm$0.02&1.05$^{+0.015}_{-0.005}$\\
$T_e$ (keV)  & 3.27$\pm$0.05& 2.67$\pm$0.01 & 2.66$\pm$0.01&2.68$\pm$0.02&2.69$\pm$0.01&2.66$\pm$0.02& 2.635$\pm$0.005& 2.61$\pm$0.01 \\
$\tau$  &3.18$\pm$0.03& 5.28$\pm$0.01 & 5.23$\pm$0.02&5.34$\pm$0.03&5.63$\pm$0.02&5.97$\pm$0.02& 6.08$\pm$0.02&6.37$\pm$0.03 \\
$R_W$ (km)& 30.61$\pm$1.55 & 38.21$\pm$ 1.75 & 39.23$\pm$1.90 & 32.95$\pm$1.55 & 29.00$\pm$1.34 & 29.69$\pm$1.45 & 27.15$\pm$1.38 & 31.35$\pm$1.61 \\
y & 0.26$\pm$0.007 &.58$\pm$0.005 &0.57$\pm$0.005& 0.60$\pm$0.008&0.67$\pm$0.005&0.74$\pm$0.008& 0.76$\pm$0.006&0.83$\pm$0.008\\
$T_{eff}$ (keV)  &1.40$\pm$ 0.01& 1.08$\pm$0.005 & 1.16$\pm$0.006&0.99$\pm$0.004&1.01$\pm$0.003&0.95$\pm$0.004&0.91$\pm$0.01& 0.86$\pm$0.01 \\
$R_{eff}(km)$& 12.92$\pm$0.61& 15.65$\pm$0.25&13.20$\pm$0.43& 18.85$\pm$0.41&23.65$\pm$0.24&26.62$\pm$0.28&27.90$\pm$0.88&29.59$\pm$0.75\\  
$F_{dbb}(\%)$ & 47.80& 20.7& 18.8& 17.43&26.90&23.36&22.39&16.56 \\
$E_l$ (keV)   &6.74$\pm$0.01& 6.73$\pm$0.06& 6.78$\pm$0.04&7.02$^{+0.04}_{-0.16}$&6.97$^{+0.05}_{-0.15}$&6.94$^{+0.08}_{-0.1}$&7.02$^{+0.3}_{-0.2}$&6.9$\pm$0.18 \\
$\sigma_l$ (keV)& 0.76$\pm$0.05 &0.80$^{+0.1}_{-0.01}$ & 0.89$\pm 0.06$&0.75$\pm$0.09&0.87$\pm$0.1&1.17$\pm$0.09& 1.21$\pm$0.2&1.45$\pm$0.16 \\
Eq. Width (eV) &207 & 227 & 241 & 154 & 161 & 222 & 205 & 295\\
$\chi^2 (d.o.f)$ & 32.4 (45) & 40.1 (45) & 54.0 (45) & 43.7 (45) & 35.5 (45) & 34.2 (45)  & 50.9 (45)   & 58.5 (45) \\
\hline
\end{tabular}
\end{minipage}
\end{table*}

\end{center}
\end{document}